# Measures of Partisan Bias
# for Legislating Fair Elections

John F. Nagle

Department of Physics

Carnegie Mellon University

Pittsburgh, PA 15213

nagle@cmu.edu

412-268-2764

**Abstract:**

Several measures of partisan bias are reviewed for single member districts with two dominant parties. These include variants of the simple bias that considers only deviation of seats from 50% at statewide 50% vote. Also included are equalization of losing votes and equalization of wasted votes, both of which apply directly when the statewide vote is not 50% and which require, not just partisan symmetry, but specific forms of the seats-votes curve. A new measure of bias is introduced, based on the geometric area between the seats-vote curve and the symmetrically inverted seats-votes curve. These measures are applied to recent Pennsylvania congressional elections and to abstract models of the seats-votes curves. The numerical values obtained from the various measures of bias are compared and contrasted. Each bias measure has merits for different seats-votes curves and for different elections, but all essentially agree for most cases when applied to measure only partisan bias, not conflated with competitiveness. This supports the inclusion of partisan fairness as a fundamental element for election law reform, and some options are discussed.

# I. Introduction

It is well recognized, in the popular press, in public opinion, and in the academic literature, that gerrymandering has been heavily practiced in redistricting, and that this is one of the likely factors in the widely perceived dysfunction of the federal government and some state governments (Mann and Ornstein 2013). While reform has occurred, notably in California and Iowa, reform should be considered in many other states, and it is therefore pertinent to consider what such reform should accomplish. Does it suffice that the redistricting process be apolitical, not involving legislators whose interests are clearly at stake?  Does it suffice that single member districts be compact and that political subdivisions be minimally subdivided?  Is it important that minorities and communities of interest obtain districts likely to elect their representatives?

In this paper, it is assumed that partisan fairness should be a primary outcome of redistricting. Conventional redistricting criteria, such as compactness and respect for political subdivisions, even when used by non-partisan map makers, may nevertheless result in political bias via unintentional gerrymandering (Grofman and King 2007, Chen and Rodden 2013). Since these criteria do not appear to suffice to bring about partisan fairness, then reform legislation should aim to create a process that does produce it.[1]  As modern computation allows the generation of many districting maps, reform could constrain the permissible pool to those that have acceptably limited political bias.[2]  Of course, the prerequisite to this kind of reform is being able to measure partisan bias.  That is the primary subject of this paper.

---

[1] This paper works within the confines of single member districts mandated by Congress for congressional districts, noting in passing that much partisan bias could be relieved by alternative voting systems (Amy 2000).

Of course, measuring partisan bias is also important for courts to decide cases of gerrymandered districting. Although gerrymandering is not justiciable in the opinions of many judges, other judges have written otherwise, but often with the reservation that a definitive measure of bias is lacking. The history has been thoroughly discussed (Grofman and King 2007). While the courts are important last resorts, partisan fairness legislation would be more expeditious, avoiding having to wait for an election and the subsequent time for a case to be decided and then wend its way through appeals. However, in the event that districting is challenged, either on partisan bias grounds or on other grounds, legislative mandating of partisan fairness would require courts to accept it as justiciable.  Furthermore, legislative guidelines for an unacceptable degree of partisan bias would also facilitate court decisions by reducing the burden of courts having to construct their own criteria *de novo*.

Partisan fairness is sometimes conflated with the concept of competitiveness, aka responsiveness or representation.  However, it has been appropriately stressed that these are separate concepts with separate measures (King and Browning 1987).  One major difference between these concepts regards how to frame their respective criteria quantitatively. Obviously, the quantifiable criterion for partisan fairness is simply to minimize partisan bias, whereas maximizing competitiveness is not so obviously a good criterion (Hirsh and Ortiz 2005, Buchler 2011). Although such considerations motivate focusing primarily on partisan fairness, quantifiable competitiveness criteria do emerge from some measures of partisan bias (McGhee 2014), so competitiveness/responsiveness will not be ignored.

---

[2] McDonald 2007. As an example, Chen and Rodden 2013 generated many maps for Florida. Although the ostensible main theme of their paper was that political geography creates political bias, there were many maps that predicted less political bias than the average randomly generated map, so reform could constrain the allowed pool to those. Mattingly and Vaughn 2014 have recently provided another notable example of generating many compactness constrained maps for North Carolina; noting how far the map used in 2012 deviated from the mean bias in their generated pool, they call for a constrained pool.

The simplest diagnostic, frequently mentioned as evidence of partisan bias, is to compare the fraction of votes won by a party with the fraction of seats won. This is a very strong diagnostic of bias when the fraction of votes won exceeds half but the fraction of seats won is less than half. However, this diagnostic is insufficient to indicate partisan bias when fractions of votes and seats are both less than half. For example, there is not necessarily partisan bias if a party wins 40% of the votes and only 20% of the seats. This could indeed be a fair outcome when many seats are highly competitive, and it conforms to the so-called "cube law" (Kendall and Stuart 1950). In the limit of all seats being totally competitive, the expected outcome is zero seats for a party that wins 1% fewer votes than its opponent; this is the winner takes all extreme of competitiveness. The assumption that the fraction of seats should equal the fraction of votes assumes that proportional representation is the ideal, and that is highly arguable (King 1989), especially as it does not allow higher levels of competitiveness. As this simple diagnostic of partisan bias is incomplete at best and also difficult to quantify generally, it is appropriate to consider more complex measures. Let it be emphasized, however, that even this simple diagnostic measure remains powerfully persuasive.[3]

Having different measures of partisan bias is somewhat similar to having different measures of compactness. Although it would be convenient if there were a single, obviously superior and universally accepted measure for any redistricting criterion, complex social problems often do not yield to simple solutions. In the case of compactness, a reasonable course has been to employ different measures and compare the results. Such a practice can also be envisioned for measuring partisan bias and the comparisons in this paper may help to promote that. However, a conclusion of this paper is that there are several measures that essentially agree, so having different measures should not block the inclusion of partisan bias in election reform.

In the next section, the well-known seats/votes (S/V) graph is reviewed as this has been a much used construct for quantitative discussion of partisan bias. Examples are given that will be used in subsequent sections to evaluate and compare measures of partisan bias; these include simple models. Also included is an apparently new way to construct the S/V graph that is applied to the 2011 Pennsylvania congressional redistricting. In section III, those examples are used to elaborate further the defect in the simple diagnostic of bias and its corresponding measure that will be designated $B_S$, where the subscript S refers to simple. Also, a much used variant, $B_{GK}$, is reviewed. A new quantitative measure of bias, $B_G$, is proposed in Section IV; it follows naturally from asymmetry in S/V graphs, being zero when the graph is symmetric.  Section V discusses several additional measures of bias that are based, in several different ways, on partisan happiness/dissatisfaction. These measures differ from the other measures in that they each lead to an ideal S/V graph and an ideal competitiveness; the $B_L$ measure, based on lost votes, leads back to proportionality and the $B_W$ measure, based on wasted votes, leads to a more competitive S/V graph (McGhee 2014).  Section VI focuses on comparing the different B measures and a more general discussion ensues in Section VII. An Appendix presents the derivation of the way that S/V curves can be obtained from election results.

---

[3] As emphasized by Baker 1990. Note that there have been many other proposed measures of partisan bias. Eight different measures were discussed by Grofman 1983.

**II.  Seats/Votes (S/V) and rank/votes (r/v) graphs**

As is well known, S/V graphs plot the predicted number of seats S versus the statewide percentage of votes V won for all districts combined. Let us begin by reviewing the bilogit family of model S/V curves.  This family has been historically important because it emphasizes the distinction between partisan bias and competitiveness by embedding two independent parameters, $\lambda$ for bias and $\rho$ for competitiveness (King and Browning 1987).  Fig. 1 shows some members of this family. The straight line for $\lambda=0$ and $\rho=1$ is the proportional S/V curve where the percentage of seats equals the percentage of the statewide vote.  The $(\lambda,\rho)=(0,3)$ curve is like the so-called cubic law; it is more responsive than proportionality, having a larger slope of 3 at V=50%, whereas the less responsive (0,0.5) curve has a smaller slope of 0.5. All curves with zero bias, $\lambda = 0$, have S=50% at V=50%, but this changes with non-zero bias.  Positive values of λ make S less than 50% when V = 50%, and by the same amount independently of the value of ρ.

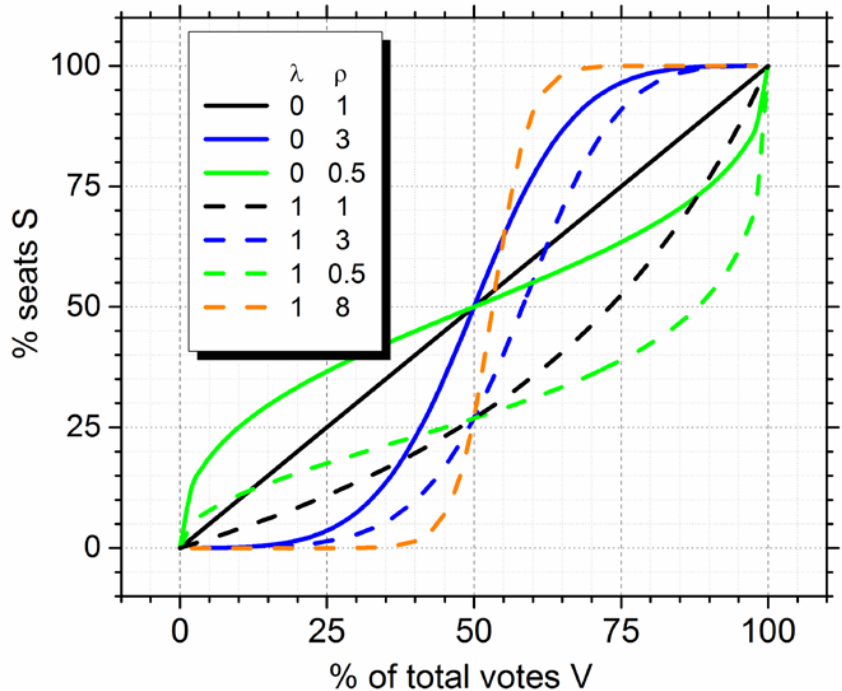

Figure 1: Bilogit S/V curves for several combinations of bias λ and responsiveness/competitiveness ρ.

While the literature clearly recognizes the importance of S/V graphs, there has been much appropriate discussion about how to construct them. Of course, to evaluate new plans before the first election, prior election returns and/or registration data by voter tabulation district (VTD) would be used.[4]

Our next example is an S/V graph for the 2011 Pennsylvania congressional districting. It is constructed using 2012 election returns.[5] The first step obtained a rank/vote (r/v) curve. To see how this was constructed, look first at the left hand red axis of Fig. 2. The districts are rank ordered from lowest %

---

[4] Backstrom, et al. 1990 advocate using election returns for a low-profile statewide base race.
[5] There were only a few 3rd party candidates and those votes were pro-rated into the percentages shown.

Republican (R) vote to highest. The R vote for each district is shown on the upper horizontal axis. Only the 12th district had a close vote; it is 6th in the rank order and was won by the Republican with 51.8% of the vote. The blue circle at 50.75% indicates the percentage of the statewide Democratic (D) vote.

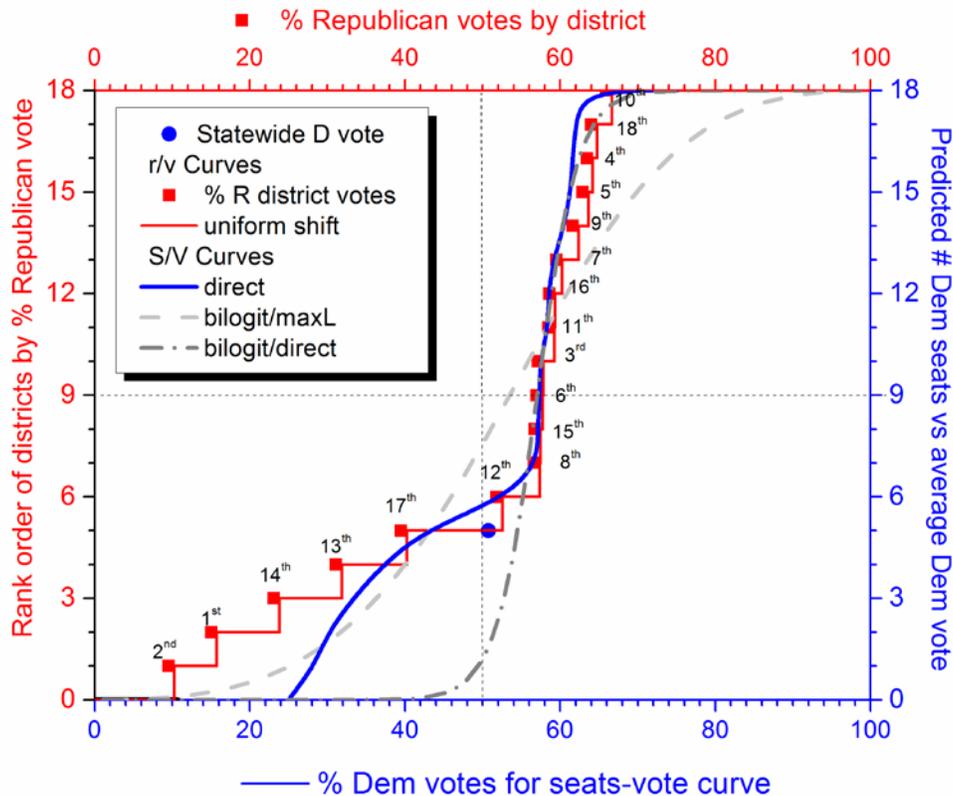

Figure 2: The red squares show the percentage Republican vote, ranked by district, in the 2012 PA congressional election. A uniform shift of 0.75% gives the Rank/Vote (r/v) curve shown as the stepped red line. The blue curve is a Seats/Votes (S/V) curve with predicted number of Democratic seats on the right hand axis versus the statewide percentage of Democratic votes on the horizontal axis. The calculation used shifts in each district proportional to the number of people in the party that must be shifted for the statewide vote to shift. The dash-dot grey curve is a direct bilogit fit to the S/V curve and the dashed grey line is the maximum likelihood bilogit approximation to the election data. The small numbers next to the red squares identify the PA districts.

If the statewide vote had been 50% for both parties, and if a uniform shift of 0.75% is made in each district, the red squares in Fig. 2 would be shifted right to the corners of the red line. That line is often taken to be the S/V curve where the number of Democratic seats S would be given on the right hand axis for the statewide Democratic vote V on the lower horizontal axis. However, this is not a valid S/V curve because it assumes that a uniform statewide swing of 40% to the Republicans would be required before they win all the seats, the last one being the one that had voted 90% Democratic in 2012. Such a uniform shift in all districts leads to the impossibility of requiring more than 100% Republicans in five of the districts.[6] The uniform shift assumption is quite unrealistic on its face because it assumes that the same number of D's would shift in districts with few D's as in districts with many D's.

A much more realistic model for obtaining an S/V curve from r/v data[7] is that a statewide percentage shift is equally likely to apply to any voter in any district; this is plausible and it has the merit that it does not allow the percentage of voters of either party in any district to fall below 0% or above 100% for any statewide vote shift. Simple math given in the Appendix[8] shows that any district, identified by subscript n, currently won by a party with $v_n$ percent of the district vote and V percent statewide vote will be lost when the statewide vote falls to $V/2v_n$. This leads to the S/V curve for PA shown in Fig. 2. Even if a district has 100% D vote when the statewide D vote is 50%, that seat would be lost when the statewide vote falls to 25%.

---

[6] The uniform shift construction of S/V curves has been appropriately criticized by many, particularly King 1989.

[7] Note that the S and the V have quite different meanings in the S/V curve, for which S is the number of expected Democratic seats S and V is the statewide Democratic vote V, than the r and v in the r/v curve, for which the r is the rank order of Republican vote and v is the district level R vote percentage.

[8] This appendix includes examples that illustrate the method and that also supplement the examples in this section.

There is little difference between the S/V and the r/v curves for the central region where 40% < V < 60%. In either case Fig. 2 predicts that Democrats would have to obtain 58% of the statewide vote to obtain half the seats. (Assuming that the rank order remained the same, that would shift districts 12, 8, 15 and 6 to the Democrats, but which particular district seats would shift is not part of an S/V curve prediction.) A swing to 58% Democrats in PA would be a landslide swing. Supposing that the mean probability of normally distributed swings is as large as 5%, there would be less than 8% probability of Democrats getting 58% or more of the total vote, so Fig. 2 predicts that Democrats would have six or fewer seats 92% of the time with the current districts. Figure 2 also emphasizes that there is little responsiveness when the vote is in the 45-55% range. In the 2014 midterm election, the overall congressional vote swung to 55% R, but there was no change in the number of party seats, nor in any of their districts. This S/V curve is even less responsive (less competitive) than proportionality; this is consistent with sweetheart bipartisan gerrymandering – incumbents of both parties like safe seats.

Another way used here to obtain an S/V curve from the PA election data employed the maximum likelihood method of King 1989. This method assumes that the underlying voting system is a member of the bilogit family. Application of this method to the r/v district data in Fig. 2 gives the values $(\lambda,\rho)=(0.32, 2.3)$.

Using these values gives the dashed grey bilogit curve shown in Fig. 2.[9]

It will also be informative to consider simpler S/V graphs for comparing measures of bias. As is well known, the best way for a party to gain advantage is to pack opponent voters into a few districts while providing a comfortable winning margin of supporters in the majority of the remaining districts. Supposing an equal number of D and R voters, an R biased districting plan could pack 25% of the districts with 80% D while providing a comfortable 60% R in 75% of the districts. This gives the S/V graph shown by the solid line in Fig. 3. The vertical jump in seats at V=60% resembles the PA S/V curve in Fig. 2. Figure 3 also shows other members belonging to a family defined by having a 75% to 25% ratio of R to D at V=50%; the corresponding composition of D is (50+3x)% in 25% of the districts and (50-x)% in the other districts. Within this family, the parameter x is roughly related inversely to the responsiveness, as there would be more shift in seats with fluctuations in V when x is small.

---

[9] Note that this curve is not the straightforward least squares fit to the S/V data points. That fit gives $(\lambda,\rho)=(2.6,9.3)$ which is shown as the dash-dot grey line in Fig. 2. It is not surprising that the bilogits fail substantially. All potential S/V curves consist of a non-zero fraction of all functions that increase monotonically with appropriate constraints on V at S=0 and the maximum S. Even if the S/V curve is only required to fit N discrete districts, the corresponding S/V family requires at least N-2 parameters as seen by considering just the finite Taylor series family consisting of N terms. Therefore, a possible S/V curve will not necessarily be well fit by the two parameter bilogit family when N>4. Rather more sophisticated methods to obtain S/V curves, at least for central ranges of V, have been proposed by Gelman and King 1994 and employed in the JudgeIt software Gelman, et al. 2012.

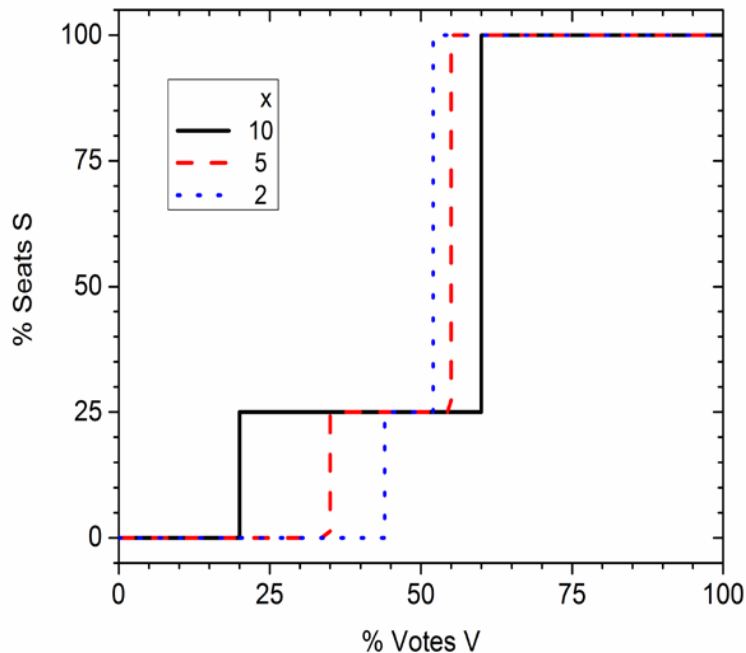

Figure 3: Members of a family of biased S/V curves that have only 25% of the seats for 50% of the votes. A first vertical jump occurs at 50-3x and a second vertical jump occurs at 50+x where the legend gives the values of x for the three curves shown.

### III. Simple bias measure $B_S$ and the $B_{GK}$ variant

S/V curves suggest a simple measure of partisan bias, designated $B_S$, based on the difference in the percentage of seats from 50% on the S/V curve when the statewide partisan vote V is 50%. For example, in Fig. 1, for all the biased cases with a bias $\lambda=1$ against the party in question, the expected percentage of seats is 26.9% when the statewide vote is 50%, so the value of $B_S$ would be set to 50% - 23.1% = 26.9%. For Fig. 2 the expected number of seats is 5 and the total number of seats is 18, so $B_S$ = (9-5)/18 which is a 22.2% bias in favor of the Republicans. (Also, in Fig. 2, for the two bilogit curves, $B_S$ = 7.9% and 42.2%, respectively.) In Fig. 3, $B_S$ = 25% for all cases. Table 1 lists the values of $B_S$ for all the S/V curves in Section II. As emphasized in the introduction, the $B_S$ measure is a persuasive indicator of partisan bias.

|   | Examples | λ | ρ | $B_S$ | $B_{GK}$ | $B_G$ | $aB_L$ | $aB_W$ |
|---|---|---|---|---|---|---|---|---|
| 1 | Bilogits | 0.5 | 3 | 12.3 | 11.9 | 7.9 | 14.2 | 4.0 |
| 2 | " | 1.0 | 3 | 23.1 | 22.6 | 15.6 | 15.1 | 7.6 |
| 3 | " | 0.5 | 8 | 12.3 | 10.3 | 3.1 | 20.3 | 8.1 |
| 4 | " | 1.0 | 8 | 23.1 | 19.9 | 6.3 | 20.4 | 8.3 |
| 5 | " maxL PA | 0.32 | 2.3 | 7.9 | 7.8 | 6.3 | 11.3 | 4.0 |
| 6 | " fit PA | 2.6 | 9.3 | 42.2 | 39.4 | 13.9 | 22.1 | 10.2 |
| --- | --- | --- | --- | --- | --- | --- | --- | --- |
| 7 | PA r/v | NA | NA | 22.2 | 20.8 | 13.7 | 13.5 | 13.9 |
| 8 | PA S/V | NA | NA | 17.8 | 17.6 | 9.4 | 17.1 | 5.3 |
| --- | --- | --- | --- | --- | --- | --- | --- | --- |
|   |   | x |   |   |   |   |   |   |
| 9 | Fig.3 | 0 | NA | NA | 0 | 0 | 0 | 25.0 | 12.5 |
| 10 | " | 1 | NA | NA | 25 | 0 | 2 | 24.1 | 11.6 |
| 11 | " | 2 | NA | NA | 25 | 2.5 | 4 | 23.1 | 10.6 |
| 12 | " | 5 | NA | NA | 25 | 25 | 10 | 20.3 | 8.1 |
| 13 | " | 10 | NA | NA | 25 | 25 | 20 | 16.3 | 10.1 |

Table 1. Rows 1-4 are bilogit S/V examples identified by the bias λ and the responsiveness ρ. The last five columns give percentage biases for the different bias measures defined in the text. Row 5 is the bilogit obtained from the PA r/v data using the maximum likelihood method and row 6 is the bilogit obtained by directly fitting to the PA S/V curve. Rows 7 and 8 are the r/v and S/V curves obtained from the PA 2012 congressional election. Rows 9-13 are for the models in Fig. 3 for the parameter values of x given in the x column.

Upon reflection, however, there is something that can go wrong with the $B_S$ measure. This is best illustrated by the examples in Fig. 3. There is definitely a bias for the x=10 case because shifts as large as 10% are rare. But by comparison, there is clearly less bias for the x=2 case because shifts exceeding 2% in the positive direction would give all the seats to the biased-against party while a 2% shift in the negative direction would not change the number of seats. As the probability of shifts can be assumed to be the same in both directions, this means that the effective bias is smaller than for the x=10% case. Indeed, partisan districters would be unlikely to regard the x=2% case as desirable and would be more likely to aim for a smaller percentage of the seats than risk losing all the seats (Owen and Grofman 1988). Note that the true bias is zero when x=0 because this is just the winner take all graph that treats both parties identically. The true bias should therefore decrease to 0 as x decreases to zero and it should do so gradually, contrary to the $B_S$ measure.

The same conclusion may be drawn from Fig. 1 for the bilogit family. In that case compare the $(\lambda,\rho)=(1,3)$ curve with the $(\lambda,\rho)=(1,8)$ curve. As noted before, both curves have the same value of $B_S$, but the argument in the previous paragraph implies that the curve with $\rho = 8$ should have less bias. Even when the bias $\lambda$ is not zero, in the limit of infinite $\rho$ the bilogit curve becomes a vertical step at V=50%, which is the unbiased, totally competitive, winner-take-all case. This further implies that the putative bias parameter $\lambda$ is not proportional to the true bias – something that we will come back to in the next section.

To overcome this deficiency in the $B_S$ measure, Gelman and King 1994 and Gelman and King 1990 have used a modified bias, to be designated here as $B_{GK}$; this first calculates the average number of seats from S(V) in the 45-55% V interval before performing the same $B_S$ calculation. Values of $B_{GK}$ are also shown in Table 1. One concern about this measure is the arbitrariness of choosing a 45%-55% interval rather than some other interval, like 47%-53%.

## IV. Geometrical measure of bias; $B_G$

Symmetry has been well recognized as indicating an absence of bias in that if one party gets 50+x% of the seats with 50+y% of the votes, then so should the other party.[10] Correspondingly, symmetry in an S/V curve means that if there is a point on the curve at 50+x% of the seats and 50+y% of the votes, then there is another point on the curve at 50-x% of the seats and 50-y% of the votes. This is called inversion symmetry about the (50,50) midpoint of the S/V graphical space. Clearly all the solid lines in Fig. 1 are symmetric, whereas none of the others are in Fig. 1 nor are any of the others in Figs. 2 and 3.

While the symmetry test diagnoses whether there is bias, one needs to actually measure it. We show in this section how to quantitatively measure the asymmetry in S/V curves geometrically. Our procedure is to first invert the original S/V curve. This maps each (S,V) point on the original S/V curve into a point at (100-S,100-V). For example, Fig. 4 shows two bilogit S/V curves and their inversions.[11] Then, we propose that bias can be measured as the geometric area between an S/V curve and its inversion. We further divide by the total number of seats so bias is given as a percentage, thereby giving a normalized value appropriate for comparing different states with different numbers of seats.

---

[10] The ratio of y to x is a measure of responsiveness. The limit for small x gives the derivative in the S/V curve at V=50%. The bilogit $\rho$ gives exactly this value when $\lambda$ is zero and $\rho$ is only 30% larger when $\lambda = 1$.

[11] If the original bilogit curve has parameters ($\lambda,\rho$), then the inverted curve is a bilogit curve with parameters ($-\lambda,\rho$) so the original bias has been reversed from one party to the other. This is a general feature of all inverted S/V curves.

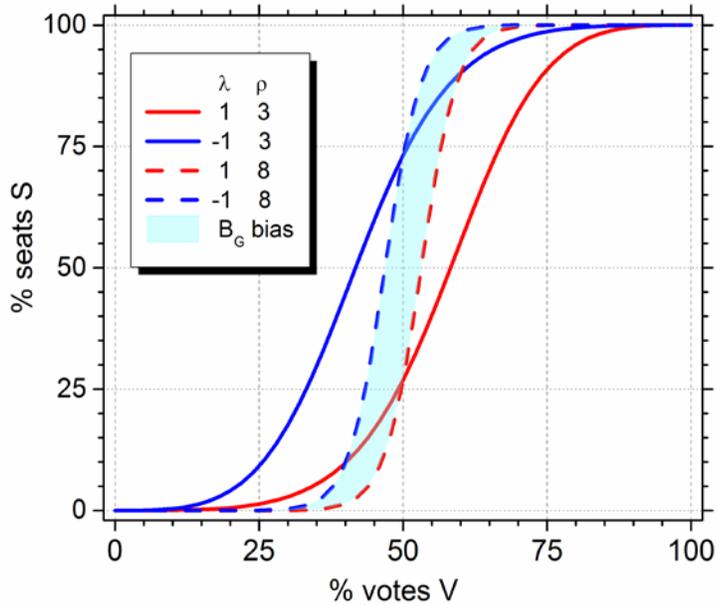

Figure 4: Inversion of bilogit S/V curves. The two solid lines are inverses of each other as are the two dashed lines. The shaded area between the two dashed lines illustrates the proposed $B_G$ measure of bias for the pair with dashed lines. The $B_G$ bias is 6.3% for this pair and 15.6% for the pair with solid lines.

Figure 5 illustrates the method for obtaining $B_G$ from the S/V curve for PA shown in Fig. 2.[12] Figure 6 shows the area of bias for one of the curves in Fig. 3. Values of $B_G$ are tabulated in Table 1.

---

[12] The computation is very simple. The %vote in the S/V Democratic data for seat number s is reassigned to seat number 18-s to obtain the S/V Republican curve. To obtain $B_G$ in a spreadsheet, one just reverses the order of the votes column, subtracts from 100%, calculates the difference in the two columns, sums the absolute values, and averages. When this procedure is applied to an r/v curve, it should first be uniformly shifted to statewide v=50%.

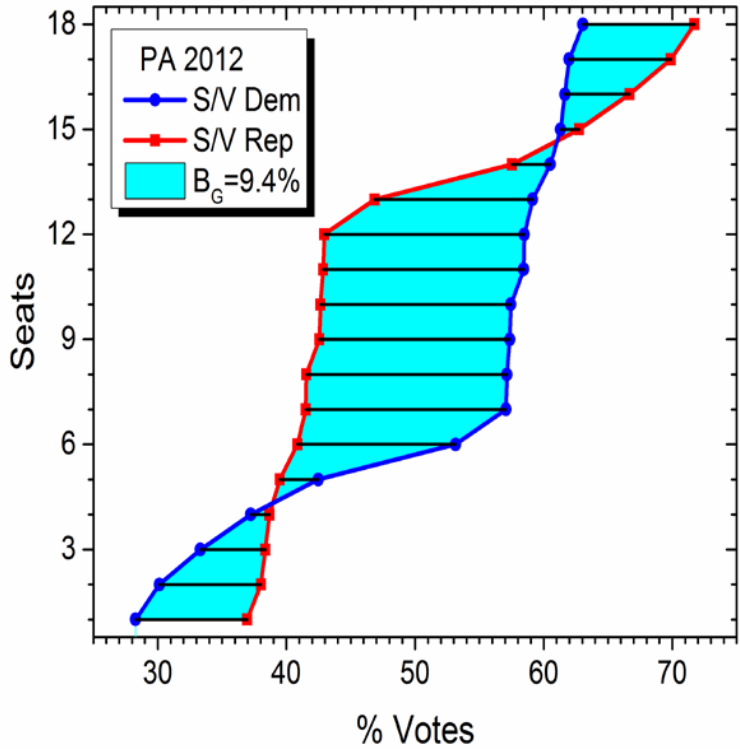

Figure 5: The S/V curve for Democrats is shown by blue squares, from Fig. 2. The red circles show the inverted S/V curve, which is the S/V curve for Republicans. The $B_G$ measure of bias is the shaded area between the two curves. It is computed to be 9.4% from the average of the length of the horizontal black lines.

Figure 6: The solid blue line shows the S/V curve for the simple model in Fig. 3 with x=10. The dashed red line is the inverted curve and the shaded area is the $B_G$ bias which equals 20%.

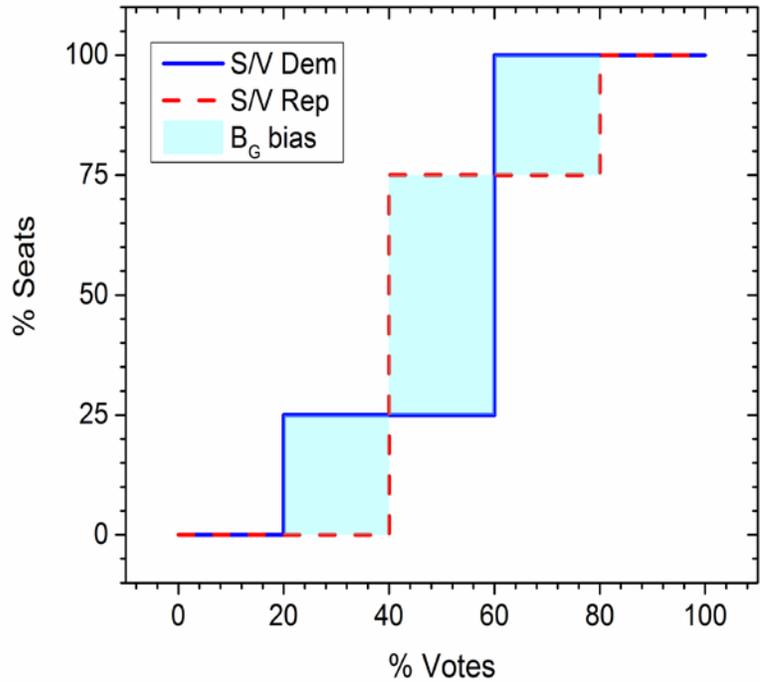

# V.  Measures based on partisan satisfaction ($B_L, B_W, B_F$)

The previous two sections focused on measures based on the S/V graph. This section considers a somewhat different approach.

As voters are happier when their candidate wins, this suggests that maximizing the total number of winning voters or minimizing the total number of losing voters would be a desirable feature of a voting system.  However, this would lead to deliberate polarization by placing all like-minded voters in the same districts.  In the limit where contiguity and compactness could be ignored, idealizing voter happiness would create districts that would be 100% likely to vote one way, so this would be a very unresponsive system, requiring an enormous swing in votes to alter the percentage of seats away from the one used in the last districting. Most reformers consider the lack of competitiveness undesirable, although the opposite view has been persuasively argued by Buchler 2011, who notes that maximizing happiness is one of the many arguments that support his position. If 100% packing could be assured, this system would be proportionately representative, but it fails in other cases. For example, if V=60% and the maximum attainable packing is 60%, then each district would have to be 60/40 and the majority party would likely win all the seats. While this is fair in that the opposite occurs when V=40%, it certainly decreases the average happiness of minority voters. As this is not desirable, and as packing like-minded voters is geographically conflicted, and as such a system is so unresponsive to changes in statewide voter preference until a subsequent redistricting, let us not further pursue total voter happiness as a goal for districting.

Nevertheless, the basic idea that a system should deal with voter happiness is appealing and leads to interesting measures of bias. The difference, compared to maximizing total voter happiness, is to use the concept in ways that directly address partisan symmetry. First, let us consider equalizing the statewide percentage of unhappy, losing voters, $L_D$ and $L_R$, in the D and R parties, respectively. The corresponding bias is then $B_L = L_D - L_R$. Interestingly, making $B_L$ zero for all values of the statewide vote $V_D$ leads to the proportional representation S/V curve. This is sufficiently important that a mathematical proof follows:

> Let $v_n$ be the percentage D vote in district n. Then $L_D$ is the sum of $v_n$ over all districts n such that $v_n$ is less than 50%. Similarly, $L_R$ is the sum of $(100 - v_n)$ over all districts n such that $v_n$ is greater than 50%. The difference $B_L = L_D - L_R$ is the sum of $v_n$ over all districts minus 100 times the number of districts won by D. The first number is just N times the percentage vote $V_D$, where N is the number of districts. The second number is just N times the percentage seats $S_D$, so $B_L = V_D - S_D$. Therefore, setting $B_L$ to zero gives the proportional S/V curve $S(V) = V$.[13] Q.E.D.

The $B_L$ measure finds bias in any deviation from proportional voting. Given an S/V curve, or even just the result of a single election, $B_L$ is just $V - S$, which is the same as the simple $B_S$ measure when $V = 50\%$.[14]

---

[13] Of course, to compare $B_L$ for different numbers of districts, one divides by N so that both S and V are represented by percentages.

[14] As equalizing losing votes leads to the popular, proportional representation, S/V curve, one might also ask what one is led to by equalizing winning votes. The same kind of analysis shows that setting the corresponding bias of winning votes to zero leads to $S(V) + V = 100\%$ which would idealize a bizarre S/V curve that would require a party to lose seats as it gains votes.

There is an important variant of the lost votes bias $B_L$ that instead minimizes the difference in so-called "wasted" votes (McGhee 2014). Wasted Democratic votes $W_D$ consist of the sum over all districts of the lost votes $L_D$ and the surplus or excess votes $E_D$, where $E_D$ is defined as the winning vote percentage in excess of 50%. A rationale for including excess votes is that they specifically focus on packing. As shown by McGhee 2014, equalizing $W_D$ and $W_R$ also leads to a particular S/V curve, namely, $S = 2V - 50\%$ when $25\% < V < 75\%$, $S = 0$ when $V < 25\%$ and $S = 1$ when $V > 75\%$. This curve is shown by the dashed red line in Fig. 7. As the slope of this curve is 2 in the midrange of V, it is more responsive than proportionality and it is more realistic for the extreme values of V.[15] Let us designate the corresponding wasted votes measure of bias $B_W$ as $W_D - W_R$. This is easily calculated as $B_W = 2V - S - 50\%$. When V=50%, similarly to $B_L$, $B_W$ is the same as the simple measure $B_S$.

One important feature of both the $B_L$ and the $B_W$ measures is that they generally have different values for different V. For example, supposing that the shifted r/v curve in Fig. 7 is the S/V curve, $B_L$ and $B_W$ are the vertical seat differences between the blue curve and the ideal $B_L=0$ and $B_W=0$ curves, respectively. Accordingly, in Fig. 7 $B_L$ indicates bias against the Democrats when statewide vote V is less than 60% but shifts to bias against the Republicans when V exceeds 60%. While this shifting complicates these measures, it also allows them to adapt when statewide votes differ considerably from 50%, as will be further discussed later. One way to obtain a different single number for these measures that is not fixated at V=50% is to average them over all V. Figure 7 shows a geometric representation of such an average. These average values for the S/V curves in Section II are what are tabulated in Table 1 under the names $aB_L$ and $aB_W$ to distinguish those numbers from the V=50% values which are the same as the $B_S$ values.

---

[15] If there are 100 seats, to obtain proportionality with 1% of the vote requires having at least half of the party's voters in one district. Generally, the slope of S/V curves must not exceed 2 when V=0. The bilogit S/V curves violate this constraint when ρ is less than 1.

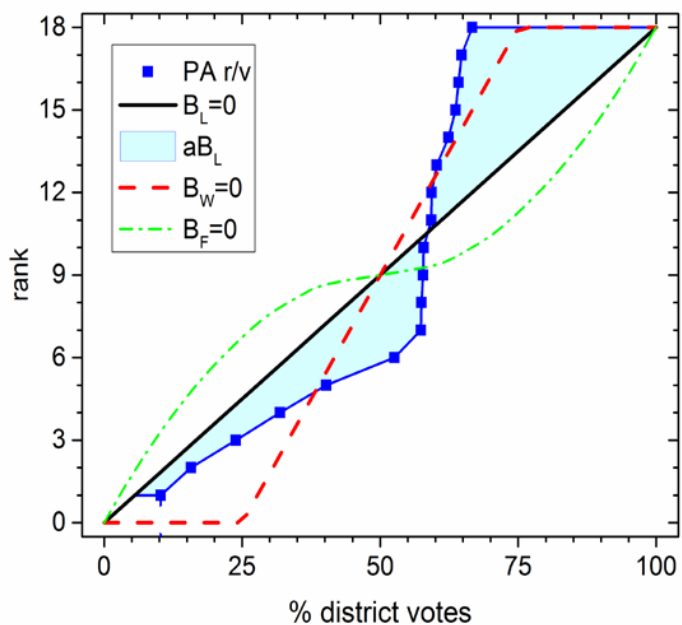

Figure 7: The PA uniform shift r/v curve compared to the solid black proportionality line. The shaded difference is the average $aB_L$ bias of the PA curve. Also shown as the dashed red line is the ideal $B_W = 0$ S/V curve. The area between it and the PA curve (not shown) is the average $aB_W$ for PA. The dot-dash $B_F = 0$ curve locates S/V coordinates when all the winning minority districts receive 50+% of the vote. $B_F = 0$ can be achieved in the region between the green dash-dot line and the proportional representation line.

Our last measure of bias returns to equalizing happy voters in both parties. Although equalizing the number of winning ("happy") votes, $H_D = H_R$, did not work, as emphasized in footnote 14, an appealing variation, at least *a priori*, is to equalize the fractions of happy voters; let us designate these fractions as $F_D$ and $F_R$. These are defined as the statewide number of winning votes for a party divided by the total number of votes for that party, so $F_D = H_D/V$ and $F_R = H_R/(1-V)$ and the corresponding measure of bias will be designated by $B_F = F_R - F_D$. The appealing feature is that an R voter and a D voter have the same probability of being happy when $B_F$ = 0. Interestingly, unlike the $B_L$ and $B_W$ measures, this one does not lead to a single ideal curve when $B_F$ = 0. Instead there are areas of the S/V plot where one can obtain $B_F$ = 0, as shown in Fig. 7. In principle, this allows some flexibility not allowed by the $B_L$ = 0 and the $B_W$ = 0 criteria, and it shares with those measures that it can be applied when V is not close to 50%. Unfortunately, it requires packing an unrealistically large fraction of the majority party voters when V is much different from 50%. For example, for V = 60%, the average fraction of majority voters in all its won districts must exceed 87%. Also, as Fig. 7 shows, the ideal $B_F$ = 0 region indulges a deeply minority party by giving it at least as many seats as proportional representation; for example, constraining $B_F$ = 0, a districting plan for a state with ten districts would ideally design a highly competitive district that would be won by the minority party with only V = 5.3% of the projected vote while giving it 10% of the seats.

## VI. Comparison of measures of bias

Let us first compare the values of $B_S$ and $B_{GK}$ in Table 1. For the bilogits and the PA S/V curves, there is not much difference because all those curves are rather smooth near V=50%, although the difference becomes larger for the larger values of the responsiveness ρ. Both measures track the bias λ used to construct the bilogit curves when a normalization factor of about 25% is applied to λ to convert it to a percentage scale.

For the S/V curves in Fig. 3, $B_{GK}$ performs somewhat better than $B_S$ in that it assigns less bias to the x=2 example. It does exhibit an artificial characteristic in that it remains 0 for all x<5/3 and stays at the maximum 25% for x>5%. This artificial piecewise behavior is caused by using an average confined to the V interval (45%,55%). This could easily be improved by using a normally distributed weighted average instead of a uniform weight requiring cutoffs that have to be arbitrarily chosen, such as at 45% and 55%.[16]

Table 1 shows that the new $B_G$ measure has the appropriate feature of increasing smoothly as x increases for the S/V curves in Fig. 3. Concomitantly, for the bilogit S/V curves, the ratio of $B_G$ for different values of responsiveness ρ differs dramatically from the $B_S$, $B_{GK}$ and λ measures, varying from about 2/3 for ρ = 3 to about 1/4 for ρ = 8. If one were to accept the $B_G$ measure as the best one, this would suggest that the bias characterizing the bilogit family be designated as λ/ρ rather than λ.

Geometrically, the difference between $B_S$, $B_{GK}$ and $B_G$ is the width of the area between the S/V plot and its inverse used in the respective calculations or the different measures of bias. $B_G$ uses all that area, $B_{GK}$ uses only the portion between 45% and 55% of the vote, and $B_S$ uses only an infinitely narrow strip at V=50%. An argument in favor of $B_G$ is that it directly uses the highly packed districts in Fig. 2, not just the few districts near V=50%.[17]

---

[16] Something equivalent appears to be embedded in JudgeIt (Gelman, et al. 2012) in that the S/V curve is calculated with statistical uncertainty.

[17] Another way to measure bias, designated here as $B_{DK}$, slices the S/V graph in the horizontal direction at the median 50% seat level and measures the the deviation from 50% vote (Singer 2015). Compared to the $B_S$ measure that slices the S/V graph in the vertical direction, $B_{DK}$ accurately treats our examples in Fig. 3. However, like the $B_S$ measure, it also may suffer in certain cases by taking too small a slice. For example, suppose half the districts have 50% R, 1/3 have 65% R and 1/6 have 20% R, so the statewide vote is 50% and the median seat also has 50% vote. Then, $B_{DK}$ would be zero whereas the expectation is for R to obtain 58% of the seats. It may be noted that the $B_G$ measure does not

As has been emphasized previously, the losing votes bias $B_L$ and the wasted votes bias $B_W$ have the same values as $B_S$ when V=50%. Since $B_L$ and $B_W$ generally vary with statewide vote V, Table 1 also shows values for their averages, $aB_L$ and $aB_W$. For the bilogit family, $aB_L$ is zero for proportional $\lambda=0, \rho=1$ representation, and it increases strongly as the responsiveness $\rho$ deviates from 1. This increase is disconcertingly stronger than with changes in $\lambda$. This pattern holds for the Fig. 3 models because 1/x is the surrogate responsiveness, suggesting that $aB_L$ is not a good measure of bias. The pattern for $aB_W$ is more complex because its ideal responsiveness is greater than proportional. $aB_W$ is generally smaller than $aB_L$ for the examples in Table 1 because those examples have responsiveness greater than proportionality $\rho = 1$. It also does not appear to be a good measure of bias.[18]

The near agreement of $aB_W$, $aB_L$ and $B_G$ for the PA S/V is not especially meaningful in view of the considerations in the preceding paragraph. Of somewhat greater interest in Table 1 is the comparison of bias for the three different PA S/V curves. The bilogit curve that is derived by maximizing statistical likelihood and the bilogit curve that is simply fit to the uniform shift S/V curve are shown in Fig. 2. These two bilogits are substantially different from each other and from the PA S/V and that is reflected by all the measures of bias in Table 1. This suggests that bilogits should not be used to evaluate real

---

slice the S/V plot in either direction but measures the entire area that involves both directions.

[18] Note that this does not imply rejection of the $B_L$ and $B_W$ measures, only the $aB_L$ and $aB_W$ measures.

districting plans.  Nevertheless, they are useful models for comparing measures of bias.

One aspect of comparing the different measures that is not adequately contained in Table 1 is the apparent advantage the $B_L$, $B_W$ and $B_F$ measures have when the statewide vote deviates substantially from V = 50%.  For example, a party with V = 70% could obtain all the seats if it can obtain a bilogit S/V curve with $\lambda = 0$ and $\rho = 8$.[19] This is a perfectly symmetrical curve that has no bias, but the high responsiveness, approaching the winner-take-all curve, gives 99.9% of the seats to the majority party at V=70%.[20] Minimizing the lost votes bias $B_L$, by achieving proportionality, would seem fairer, but this becomes difficult to implement when V increases further. As noted before, a similar problem arises for $B_F$. This problem is avoided by choosing instead to minimize the wasted votes bias $B_W$ and this has the advantage to those concerned more with competitiveness of being more responsive to swings in the statewide vote. However, this measure would freeze out parties that have less than 25% of the vote. While these are very real considerations, it must be emphasized that these issues are concerned with competitiveness/responsiveness, not partisan bias. For partisan bias, the $B_L$ and $B_W$ measures should be applied only to the V = 50% point on the S/V curve, at which they agree with the simple $B_S$ measure. From this point of view, applying the $B_L$ or $B_W$ measures when V is not equal to 50% attempts to do too much in addition to assessing partisan bias.

It may finally be noted that Table 1 does not contain values of $B_F$ because it is not possible to calculate it for the bilogits or for the Fig. 3 curves because those are only S/V curves without detailed voter information. It is easy to calculate $B_F$ from the PA data in Fig. 2. For that election only 45% of the D voters voted for a winning candidate while 87% of the R voters did. Subtracting these yields a bias $B_F$ = 45% favoring the Republicans. In percentage terms, this is the most sensitive measure of bias for PA.[21]

---

[19] This curve is halfway between the two $\rho$ = 8 curves in Fig. 4.

[20] Something like this might account for Maryland having no Republicans in Congress.

   Let us also compare how measures of bias change when using the uniformly shifted PA r/v curve instead of the S/V curve. This is of interest because r/v curves are directly produced from data, whereas the S/V curve requires a model and a layer of intervening manipulation. Discounting the $aB_L$ and $aB_W$ measures for the reasons discussed above, Table 1 shows that only modestly larger values for the $B_S$, $B_{GK}$, and $B_G$ measures of bias are obtained for the r/v curve.[22] This suggests that measures of bias could be based on r/v curves, but it must be emphasized that this is only valid if the r/v curve is first shifted to v=50% statewide vote as is emphasized in Fig. 9 in the Appendix; that is a manipulation that is not made when constructing the S/V curve, albeit a simpler one than those that are used in constructing S/V curves.

---

[21] One might prefer to describe these results by saying that twice as many R voters in PA were happy with their representative sent to Congress as were D voters. It may also be noted that these values were computed from the r/v data with no uniform shift. They change only by about 0.2% when the uniform shift was applied because no seats shifted. However, as uniform shift is increased, seats shift and $B_F$ decreases. An 8% shift would flip 5 seats (giving a Democratic majority of 10 to 8) and then $B_F$ reverses sign and becomes an 8% bias favoring Democrats.

[22] The ratio for $B_G$ is largest as it gives more emphasis to districts won by large percentages and those are most affected by the S/V model calculation.

**VII. Discussion**

There are several aspects regarding inclusion of partisan fairness in the law. The first aspect distinguishes between making it justiciable, in order to challenge blatantly biased districting plans on the one hand and, on the other hand, including it in legislation and/or state constitutional amendments in order to avoid having biased plans in the first place. In either case, conciseness would be helpful. As Grofman and King 2007 have focused on the judiciary, this discussion concentrates on legislation. Of course, to reform existing election law, it is most important that partisan bias simply be recognized as a restrictive criterion in state constitutions. This could be done just with a phrase in a constitutional amendment. Unfortunately, that could allow it to be practically disregarded just as compactness has been largely disregarded in the past, so something more might be contemplated. Reform could also come through legislative statute; that has the advantage of being easier to change as conditions change, but also easier to weaken for partisan advantage. Reform could also come through an independent districting commission. Such a commission would likely be more inclined to include partisan fairness in its deliberations if there is at least a phrase in the constitution mandating it, and it would be largely prohibited from doing so if political data, like election returns or voter registration, were constitutionally forbidden, as some reformers advocate.[23] An independent commission might also find it helpful if a clear set of guidelines is worked out by scholars.

---

[23] However, it may be noted that Belin, et al. 2011 propose an interesting way to promote partisan fairness that does not require political data, only population density data.

The second aspect regards how measures of bias should be applied. In their important paper Grofman and King 2007 discuss five possibilities: (1) "Require plans with as little partisan bias as practicable", (2) "Disqualify plans with partisan bias that deviate from symmetry by at least one seat", (3) "Disqualify only those plans with egregious levels of partisan bias (defined in terms of a specified percentage threshold)", (4) "Disqualify only those plans that (can be expected to) translate a minority of the votes into a majority of the seats", (5) "Disqualify only those plans whose partisan bias is both severe and greater than that in the plan being replaced." Although Grofman and King 2007 discussed these in the context of justiciability, these are also possibilities for the legislative approach. All except number (4) require a quantifiable measure of bias. Possibilities (2) and (3) also address a third aspect, namely, how much partisan bias should be allowed.

The fourth aspect of including partisan fairness regards measures of bias to be used.[24] That has been the focus of this paper, and several different measures have been presented and compared to each other. Recapitulating, when applied appropriately to the S/V or the shifted r/v curves, all the viable measures essentially agree.[25] Apparent divergences that occur when the $B_L$ (proportional representation) and the $B_W$ (super-proportional representation) are applied to statewide votes that deviate considerably from 50% are due to these measures also attempting to establish a standard for competitiveness. As noted in the introduction, that is not an easy issue; it should not impede going forward with the issue of partisan fairness. In the remainder of this section, the different bias measures will be discussed in relation to the other three aspects of inclusion of partisan fairness in the law.

---

[24] Grofman and King 2007 essentially assume that bias would be measured using JudgeIt software (Gelman, et al. 2012). This would be a good choice, although it would probably be considered inappropriate to name it in legislation. It may also be noted that JudgeIt is apparently not designed to reliably obtain the S/V curve over the full range of V from 0 to 100%, so it is incompatible with the $B_G$ measure. As mentioned at the end of Section VI, it might be simpler just to use the shifted election data for measuring partisan bias, and JudgeIt does seem to involve a uniform shift. Of course, JudgeIt is an invaluable statistical tool for measuring many other election elements and for including other kinds of data, especially incumbency. In this latter regard, it may also be noted that the new incumbent in the $12^{th}$ PA district in 2014 received 3.8% more than the average R vote in 2014 and 2.6% more in 2012, consistent with a 1.2% incumbency advantage based on a very small sample.

    The simple $B_S$ measure has the merit that it can be more concisely stated than the others. That is an advantage when one considers that any measure of bias would already have to involve quite a lot of non-trivial language if it is to be spelled out in detail. First, it would have to be stated what the measure would be applied to. A guideline for a commission could give the choice between an S/V curve or a shifted r/v curve, emphasizing that the measure would be applied to V = 50% in either case. Legislative statute would most likely have to make the choice between S/V and r/v. Second, there is the consideration of what kind of data to use in drawing the S/V or r/v curve (see footnote 4), again with similar distinctions regarding legislation or redistricting commissions.

    Regarding the second aspect of how to apply a measure of bias for possibilities (1,2,3,5, listed above) of Grofman and King 2007, none of the viable measures is any less preferred than the others. Also regarding the third aspect of how much

---

[25] We do not include the unviable $aB_L$ and the $aB_W$ measures. The $B_G$ measure can be made to agree better with the others if it is multiplied by a factor of 1.5.

bias to allow, the only distinction is that any percentage allowed by $B_G$ should be set smaller than the percentages allowed by $B_S$, $B_{GK}$, $B_L$ and $B_W$, and the percentages for $B_F$ should be set higher.[26] Possibility (2) of Grofman and King 2007 above requires that the allowable bias would be smaller for a body with more seats. For PA with 18 congressional seats, it would constrain bias to 5.5% using the $B_S$, $B_{GK}$, $B_L$ and $B_W$ measures and to about 4% using the $B_G$ measure.[27]

    A final issue then is whether a percentage bias number should be embedded in the state constitution or whether it is best done by legislative statute. While reform would be more robust if it could be embedded in the constitution, the language to cover all the contingencies could be unwieldy and, to ensure passage, the bias range may be set too high and then not be easy to change. This consideration favors possibility (1) above, but the clause "as practicable" could vitiate implementation if it were claimed that other factors make fairness not practicable. As usual, reform law would require political deliberation, but that should not dissuade reformers from pressing the issue of partisan fairness in election law.

---

[26] Different levels for different measures is analogous to obtaining different numbers when measuring the same length with a meter stick or a yard stick, so this is not a valid criticism against the quantification of partisan bias.

[27] One concern could be that political geography would not allow any plans to fall in the allowed bias range. This occurred with the districts generated by Chen and Rodden 2013 for Florida. For PA that concern might be moot as it has been asserted that a districting plan has been made that would reverse the 2012 result, giving the Democrats 13 seats if all voters would have voted for candidates of the same party (Leach 2014).

**Appendix: Construction of S/V curves from voting data.**

Given N districts, labeled n=1,…,N, and given Democratic (D) past (or projected) vote $v_{nD}$ in current (or proposed) district n, the statewide D vote is $V_D = (1/N)\Sigma v_{nD}$ where $\Sigma$ connotes the sum over all districts. First consider a statewide shift from D to R (Republicans) and denote the new statewide vote as $V_D' = V_D - \delta_D$, where $\delta_D$ is the average percentage of shifted Democrats statewide. We assume that there is an equal probability $x_D$ for any Democrat in any district to shift,[28] so the new district vote is $v_{nD}' = v_{nD}(1-x_D)$. Summing $v_{nD}'$ over all districts gives $V_D' = V_D - x_D V_D$; this identifies $x_D = \delta_D/V_D$. Then, the shift $\delta_{nD}$ in statewide vote that is necessary to shift a seat n that was won by the Democrats must satisfy $v_{nD}' = 50\%$ which gives $\delta_{nD} = + V_D - V_D/2v_{nD}$. Therefore, the statewide vote required to shift the nth seat from D to R is $V_{nD}' = V_D - \delta_{nD} = V_D/2v_{nD}$, as stated in the main text. This is the value of V that is plotted on the S/V curve for the nth seat originally won by the Democrats. For seats originally won by the Republicans, the same derivation gives the symmetrically equivalent formula $V_{nR}' = V_R/2v_{nR}$ when the Republican vote $v_{nR}$ exceeds 50%. For those seats, the value of V that is plotted on the $S_D/V_D$ curve for district n is, of course, $100 - V_{nR}'$, corresponding to $V_{nD}'$.

It may be of some interest to view some simple r/v examples and their corresponding S/V curves. Figure 8 shows that uniform distributions of the votes by ranked seats result in S/V curves that all have $B_S = 0$. Uniform shifts of these r/v lines to V =

---

[28] The model assumes that only Democrats shift when there is a statewide shift towards Republicans. Supposing that Republicans also shift with a uniform probability leads to the unreasonable result that a massively Republican district would always become more Democratic even when the statewide vote was shifting in favor of the Republicans.

50% (not shown) are all symmetrical. Slight asymmetry is evident for the two S/V examples derived from r/v curves with statewide vote different from 50%, so the $B_G$ measure is small but not zero, gradually reaching 5% as V increases to 75%.

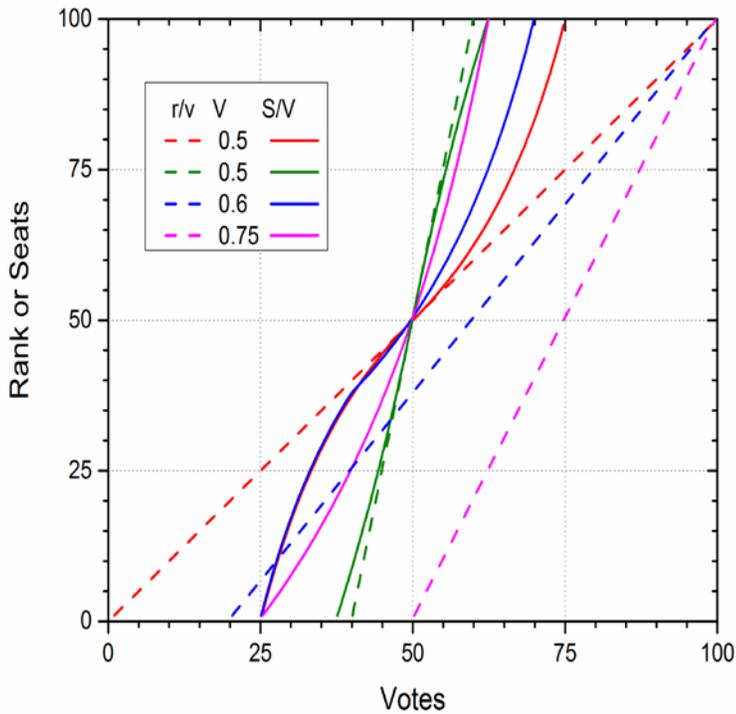

Figure 8. Four simple r/v examples are shown by the dashed lines and the corresponding S/V curves are shown by solid lines of the same color. For the r/v lines the votes axis is the vote for the first party for each ranked seat. For the S/V curves the votes are the statewide votes required for the second party to obtain the number of seats on the S/V curve. For the last example, the first party received 75% of the statewide vote, which is the average of the vote in 100 districts starting with 50% for the seat won narrowly and moving up to the last seat which was won unanimously. All of these examples are unbiased using the $B_S$ measure.

In contrast to Fig. 8, Fig. 9 shows an example of an asymmetrical r/v curve where the first party wins only 25% of the seats with 37.5% of the vote. Nevertheless, this outcome is biased against the second party because the S/V curve shows that the second party would, on average, only win 38% of the seats when the statewide vote is 50% for a bias $B_S$ = 12%.

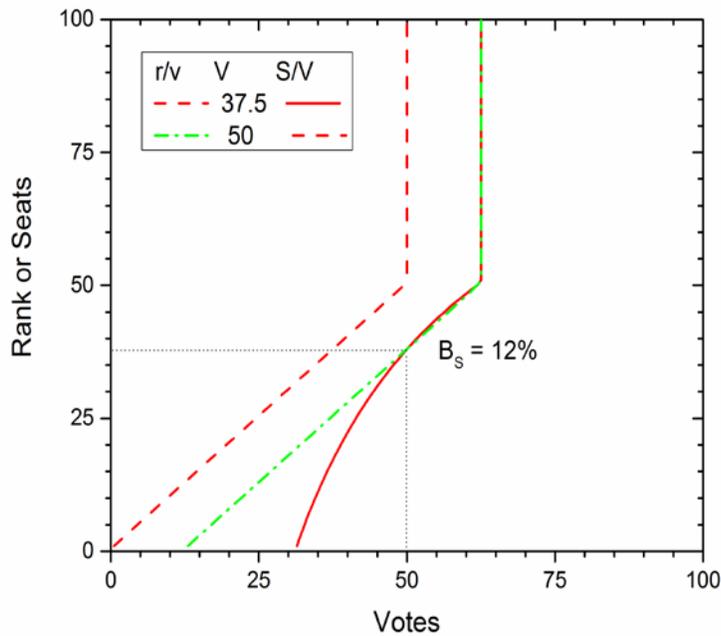

Figure 9. A highly asymmetric r/v curve that gives 37.5% of the statewide vote to the first party is shown by the red dashed line. That party loses ranked seats 1-50 and splits seats 51-100. The green dash-dot line shows the shifted r/v curve to 50% statewide vote and the solid red curve shows the S/V curve computed from the unshifted r/v curve. The S/V and the shifted r/v curves agree in the vote region centered around 50%.

Finally, it may be of interest to view a comparison of the S/V curves for the 2014 and 2012 PA congressional elections. Fig. 10 shows that three uncontested seats moved the largest and smallest district vote percentages $v_n$ to more extreme values. More interestingly, it also appears that the "wall" of Republican seats built up near $V_D$ = 60% in 2012 may be moving into the more competitive range, even though those seats were comfortably held by the Republicans at the 60% district level in 2014 due to a 5.5% statewide swing. This would be consistent with studies that have reported gradual erosion in the effects of gerrymanders (McGhee 2014). This should not, of course, be used as an excuse to ignore gerrymandered partisan bias when redistricting, even if the effect only lasts for a few elections.

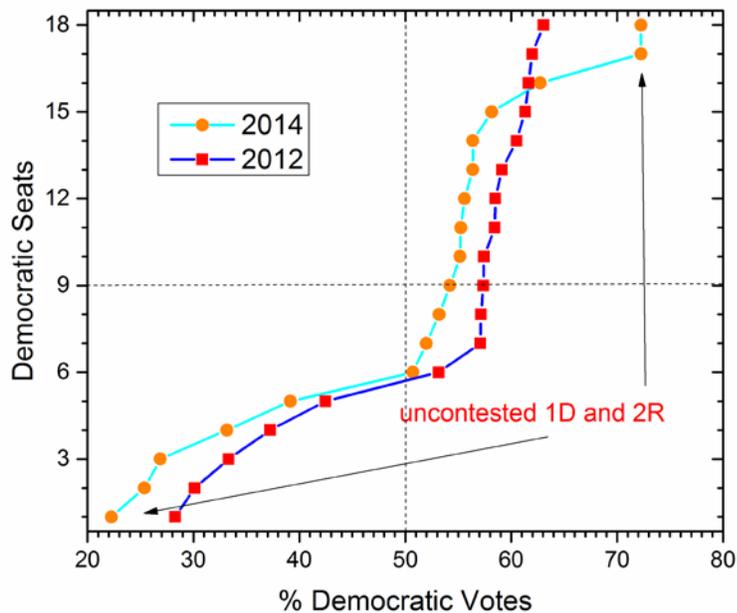

Figure 10. Comparison of S/V curves for PA 2012 and 2014 Congressional elections. Three uncontested seats, one Democratic and two Republican, are at the ends of the 2014 curve.

## Bibliography

Amy, Douglas J.  2000.  *Behind the Ballot Box: A Citizen's Guide to Voting Systems*.  Westport, CT: Praeger Publishers.
Backstrom, C, L Robins, and S Eller.  1990.  "Establishing a Statewide Electoral Effects Baseline."  In *Political Gerrymandering and the Courts*, ed.  B. Grofman.  New York, NY: Agathon Press.  145-70.
Baker, G. E.  1990.  "The "Totality of Circumstances" Approach."  In *Political Gerrymandering and the Courts*, ed.  B. Grofman.  New York, NY: Agathon Press.  203-11.
Belin, T. R., H. J. Fischer, and C. M. Zigler.  2011.  "Using a Density-Variation/Compactness Measure to Evaluate Redistricting Plans for Partisan Bias and Electoral Responsiveness."  *Statistics, Politics, and Policy* 2.
Buchler, Justin.  2011.  *Hiring and Firing Public Officials: Rethinking the Purpose of Elections*.  New York, NY: Oxford University Press.
Chen, J. W., and J. Rodden.  2013.  "Unintentional Gerrymandering: Political Geography and Electoral Bias in Legislatures."  *Quarterly Journal of Political Science* 8: 239-69.
Gelman, A., and G. King.  1990.  "Estimating the Electoral Consequences of Legislative Redistricting."  *Journal of the American Statistical Association* 85: 274-82.
———.  1994.  "A Unified Method of Evaluating Electoral Systems and Redistricting Plans."  *American Journal of Political Science* 38: 514-54.

Gelman, A., G. King, and A. C. Thomas. 2012. "Judgeit Ii: A Program for Evaluating Electoral Systems and Redistricting Plans." *http://gking.harvard.edu/judgeit*.

Grofman, B., and G. King. 2007. "The Future of Partisan Symmetry as a Judicial Test for Partisan Gerrymandering after Lulac V. Perry." *Election Law Journal* 6: 2-35.

Grofman, Bernard. 1983. "Measures of Bias and Proportionality in Seats-Votes Relationships." *Political Methodology* 9: 295-327.

Hirsh, S., and D. Ortiz. 2005. "Beyond Party Lines: Principles for Redistricting Reform." http://www2.illinois.gov/gov/reformnow/Documents/press%20releases/Principles%20for%20redistricting%20reform.pdf. (2005).

Kendall, M. G., and A. Stuart. 1950. "The Law of Cubic Proportion in Election Results." *British Journal of Sociology* 1: 183-97.

King, G. 1989. "Representation through Legislative Redistricting - a Stochastic-Model." *American Journal of Political Science* 33: 787-824.

King, G., and R. X. Browning. 1987. "Democratic Representation and Partisan Bias in Congressional Elections." *American Political Science Review* 81: 1251-73.

Leach, Daylin. "Dismantling Democracy: How Gerrymandering Has Stolen Your Voice and Undermined Our Political System." http://www.roxburynews.com/index.php?a=6830

Mann, T. E., and N. J. Ornstein. 2013. "Finding the Common Good in an Era of Dysfunctional Governance." *Daedalus* 142: 15-24.

Mattingly, J. C., and C. Vaughn. 2014. "Redistricting and the Will of the People." *arXiv* 1410.8796v1 [physics.soc-ph] 29.

McDonald, M. P. 2007. "Regulating Redistricting." *Ps-Political Science & Politics* 40: 675-79.

McGhee, E. 2014. "Measuring Partisan Bias in Single-Member District Electoral Systems." *Legislative Studies Quarterly* 39: 55-85.

Owen, G, and B.. Grofman. 1988. "Optimal Partisan Gerrymandering." *Political Geography Quarterly* 7: 5-22.

Singer, Jeff. "Just How Strong Are the Gop's Gerrymanders? Daily Kos Elections' Median District Scores Explain." http://www.dailykos.com/story/2015/03/19/1370484/-Just-how-strong-are-the-GOP-s-gerrymanders-Daily-Kos-Elections-median-district-scores-explain?detail=email. .